\documentclass[10pt,a4paper,dvipsnames,usenames]{article}
\usepackage{amssymb}
\usepackage{amsmath}
\usepackage{amsthm}
\usepackage{latexsym}
\usepackage[dvips]{epsfig}
\usepackage{mathrsfs}
\usepackage{bm}
\usepackage{stmaryrd}
\usepackage{authblk}

\usepackage[dvipsnames]{xcolor}
\usepackage{tikz}
\usepackage{enumitem}
\usepackage{commath}
\usepackage{tipa}
\usepackage{newtxtext}
\usepackage{stmaryrd}

\theoremstyle{plain}
\newtheorem{proposition}{Proposition}
\newtheorem{lemma}{Lemma}

\newtheorem{remark}{Remark}

\allowdisplaybreaks

\def\bma{{\bm a}}
\def\bmb{{\bm b}}
\def\bmc{{\bm c}}
\def\bmd{{\bm d}}
\def\bme{{\bm e}}

\def\bmg{{\bm g}}
\def\bmh{{\bm h}}
\def\bmi{{\bm i}}
\def\bmj{{\bm j}}
\def\bmk{{\bm k}}

\def\bmo{{\bm o}}
\def\bmq{{\bm q}}

\def\bmt{{\bm t}}

\def\bmzero{{\bm 0}}
\def\bmone{{\bm 1}}

\def\bmA{{\bm A}}
\def\bmB{{\bm B}}

\def\bmK{{\bm K}}

\def\bmepsilon{{\bm \epsilon}}
\def\bmeta{{\bm \eta}}

\def\bmiota{{\bm \iota}}
\def\bmomega{{\bm \omega}}

\def\bmsigma{{\bm \sigma}}

\def\bmpartial{{\bm \partial}}
\def\bmnabla{{\bm \nabla}}

\usepackage{mathtools}% http://ctan.org/pkg/mathtools
\usepackage{xparse}% http://ctan.org/pkg/xparse
\makeatletter
\newcommand{\raisemath}[1]{\mathpalette{\raisem@th{#1}}}
\newcommand{\raisem@th}[3]{\raisebox{#1}{$#2#3$}}
\makeatother
\NewDocumentCommand{\newrbar}{O{0pt} O{0pt}}{
  \ensuremath{\mathrlap{\raisemath{#2}{\hspace*{#1}{\mathchar'26\mkern-9mu}}}r}}

\newcounter{mnotecount}%[section]

\newcommand{\mnotex}[1]%{}
{\protect{\stepcounter{mnotecount}}$^{\mbox{\footnotesize $\bullet$\themnotecount}}$ 
\marginpar{%\color{red}%
\raggedright\tiny\em
$\!\!\!\!\!\!\,\bullet$\themnotecount: #1} }

\newcounter{mnote}

\usepackage{color}
\usepackage{xcolor}
\usepackage[normalem]{ulem}
\usepackage{soul}
\usepackage{hyperref}

\begin{document}

\title{\textbf{Calculation of asymptotic charges at the critical sets of null infinity}}
 
\author[1]{ Mariem Magdy Ali Mohamed \footnote{E-mail
    address:{\tt m.m.a.mohamed@qmul.ac.uk}}}
%\author[1]{Con T. Ributor}

\affil[1]{School of Mathematical Sciences, Queen Mary, University of London,
Mile End Road, London E1 4NS, United Kingdom.}

\maketitle

\begin{abstract}
The asymptotic structure of null and spatial infinities of asymptotically flat spacetimes plays an essential role in discussing gravitational radiation, gravitational memory effect, and conserved quantities in General Relativity. Bondi, Metzner and Sachs established that the asymptotic symmetry group for asymptotically simple spacetimes is the infinite-dimensional BMS group. Given that null infinity is divided into two sets: past null infinity $\mathscr{I}^{-}$ and future null infinity $\mathscr{I}^{+}$, one can identify two independent symmetry groups: $\text{BMS}^{-}$ at $\mathscr{I}^{-}$ and $\text{BMS}^{+}$ at $\mathscr{I}^{+}$. Associated with these symmetries are the so-called BMS charges. A recent conjecture by Strominger suggests that the generators of $\text{BMS}^{-}$ and $\text{BMS}^{+}$ and their associated charges are related via an antipodal reflection map near spatial infinity. To verify this matching, an analysis of the gravitational field near spatial infinity is required. This task is complicated due to the singular nature of spatial infinity for spacetimes with non-vanishing ADM mass. Different frameworks have been introduced in the literature to address this singularity, e.g., Friedrich's cylinder, Ashtekar-Hansen's hyperboloid and Ashtekar-Romano's asymptote at spatial infinity. This paper reviews the role of Friedrich's formulation of spatial infinity in the investigation of the matching of the spin-2 charges on Minkowski spacetime and in the full GR setting.
\end{abstract}

\tableofcontents

\newpage

\section{Introduction}
In classical General Relativity (GR), isolated systems are commonly described by asymptotically flat spacetimes, with a metric approaching the Minkowski metric far from the source. In this setting, the influential work of R.~Penrose \cite{Penrose63,Penrose64} offers a geometrical approach to the studies of isolated systems. In particular, Penrose's notion of "asymptotic simplicity" identifies spacetimes with a conformal extension similar to that of Minkowski spacetime, implying the existence of null infinity $\mathscr{I}$, comprised of two disjoint sets: future null infinity $\mathscr{I}^{+}$ and past null infinity $\mathscr{I}^{-}$. The universal fields shared by asymptotically simple spacetimes allow us to identify the infinite-dimensional asymptotic symmetry group known as the BMS group \cite{BMS62}, named after Bondi, Metzner and Sachs. However, Penrose's notion of asymptotic simplicity is not concerned with the behaviour of the gravitational field at spatial infinity, which is a crucial ingredient in the discussion of conserved quantities in GR \cite{Geroch77}.

One of the challenging aspects of studies of the asymptotic structure at spatial infinity is the singular conformal structure at spatial infinity $i^{0}$ for spacetimes with non-vanishing Arnowitt-Deser-Misner (ADM) mass. Different formulations of spatial infinity \cite{AshHan78,AshRom92,Friedrich98} can be used to resolve the structure of the gravitational field in this region. Of particular importance to this article is Friedrich's formulation of spatial infinity, initially introduced in \cite{Friedrich98} with the goal of obtaining a regular initial value problem at spatial infinity for the so-called conformal Einstien field equations. This representation of spatial infinity is linked to the conformal properties of spacetimes, and it introduces a blow-up of the spatial infinity point $i^{0}$ to a cylinder $(-1,1) \times \mathbb{S}^2$ commonly known as the cylinder at spatial infinity $\mathcal{I}$. The cylinder $\mathcal{I}$ touches the endpoints of past and future null infinities $\mathscr{I}^{\pm}$ at the critical sets $\mathcal{I}^{\pm} = \{ \pm 1 \} \times \mathbb{S}^2$. This representation of spatial infinity is useful for relating quantities at the critical sets $\mathcal{I}^{\pm}$ to initial data on a Cauchy hypersurface ---- see, e.g, \cite{FriedrichKannar00, GasperinKroon20}. Other equally significant contributions to the studies of the asymptotic structure at spatial infinity are Ashtekar-Hansen's and Ashtekar-Romano's formulations of spatial infinity --- see \cite{AshHan78,AshRom92}. While the relation between Friedrich's and Ashtekar-Romano's formulations was established in \cite{MohamedKroon21}, the link between Friedrich's formulation and Ashtekar-Hansen's remains unexplored. Ashtekar-Hansen's and Ashtekar-Romano's formulations introduce the Spi group, denoting the infinite-dimensional asymptotic symmetry group at spatial infinity, with a structure similar to the BMS group at null infinity. 

One physical motivation for studying symmetries is the prospect of defining conserved quantities (also known as Noether charges or simply charges) in an isolated system, e.g., energy, momentum and angular momentum. For fields on a fixed background, conserved quantities are defined by considering the integral of the local energy-momentum tensor contracted with Killing fields of the background spacetime over a Cauchy hypersurface. Given the dual role of the spacetime metric in GR, describing both geometrical and physical aspects of the theory, one can only define such conserved quantities in the asymptotic limit, where the background and physical fields can be studied separately --- see \cite{Geroch72}. As noted in \cite{WaldZoupas00}, there had been a clear distinction in earlier work in defining conserved quantities associated with asymptotic symmetries at null and spatial infinity. In particular, the Hamiltonian formulation was prominently used in the derivation of conserved quantities at spatial infinity \cite{ADM59,Witten62,ReggeTeitelboim74} compared to null infinity. The challenge in defining "conserved quantities" at null infinity using a standard Hamiltonian formulation is that symplectic current can be radiated away at null infinity, and thus, generically, there exists no Hamiltonian generating BMS transformations. However, the general prescription in \cite{WaldZoupas00} allows one to define charges associated with asymptotic symmetries even in situations where the Hamiltonian does not exist --- see also \cite{DrayStreubel84,GodPerry20}. This prescription was used in \cite{GrantPrabhuShehzad22} to derive explicit expressions for the charges associated with BMS symmetries at null infinity. Finally, note that "conserved quantities" at null infinity are not exactly conserved for general dynamical spacetimes. Instead, a charge associated with a BMS symmetry will have a non-vanishing flux through null infinity $\mathscr{I}^{\pm}$ \cite{WaldZoupas00}. 

The discussion in this article is motivated by a recent conjecture by Strominger \cite{Strominger14} suggesting that BMS symmetries and their associated charges can be linked to soft theorems \cite{Weinberg65,Strominger14,HLMS15} and the gravitational memory effect \cite{Favata10,Christodoulou91,BD92}. This link is based on the so-called matching problem, i.e., the idea that the BMS groups at past and future null infinities ($\text{BMS}^{-}$ at $\mathscr{I}^{-}$ and $\text{BMS}^{+}$ at $\mathscr{I}^{+}$) can be matched by an antipodal reflection map near spatial infinity. The matching of these symmetries leads to a global diagonal asymptotic symmetry group in GR, implying that the incoming flux of an asymptotic charge at $\mathscr{I}^{-}$ would be equal to the outgoing flux of the corresponding charge at $\mathscr{I}^{+}$. Generically, the matching of $\text{BMS}^{+}$ and $\text{BMS}^{-}$ and their associated charges requires an analysis of the gravitational field and the charges near spatial infinity. One significant challenge in this analysis is the singular nature of the conformal structure at spatial infinity $i^{0}$, further highlighting the importance of the different representations of spatial infinity \cite{AshHan78,BeigSchmidt82,AshRom92,Friedrich98} in the discussion of the matching problem. In recent years, numerous articles discussed the asymptotic symmetry group at spatial infinity \cite{CampEyh17,HenneauxTroessaert18,HenneauxTroessaert18-2,HenneauxTroessaert18-3,HenneauxTroessaert19-2,PrabhuShehzad20} and their matching with the asymptotic charges at null infinities \cite{CampEyh17,Troessaert18,Prabhu18,Prabhu19,PrabhuShehzad22,CaponeNguyenParisini23}. On Minkowski spacetime, the matching of supertranslation asymptotic charges has been investigated for the spin-1 field in \cite{CampEyh17} and the spin-2 field in \cite{Troessaert18}. For spacetimes satisfying Ashtekar-Hansen's definition of asymptotic flatness at null and spatial infinity (see \cite{AshHan78} or \cite{Prabhu19} for precise definition), the matching of supertranslation asymptotic charges for the spin-1 and the gravitational field was shown in \cite{Prabhu18,Prabhu19}. Moreover, the matching for Lorentz charges was also investigated in \cite{PrabhuShehzad22} using Ashtekar-Hansen's formulation of spatial infinity.

The purpose of this paper is to provide a streamlined presentation of the calculation of asymptotic charges at the critical sets and their matching using Friedrich's formulation of spatial infinity. The expressions of the asymptotic charges used in this article are adapted from \cite{Prabhu19}, which agrees with the general prescription of conserved quantities given in \cite{WaldZoupas00}. The use of Friedrich's formulation allows us to express the supertranslation asymptotic charges at $\mathcal{I}^{\pm}$ in terms of initial data given on a Cauchy hypersurface, and to show that the matching of the asymptotic charges at $\mathcal{I}^{\pm}$ follows from certain regularity conditions on the free initial data. The full analysis of the asymptotic charges in a full GR setting using Friedrich's formulation will be presented elsewhere. However, the main results can be summarised as follows:
\begin{center}
    \emph{For the generic initial data set given in \cite{Huang10}, the asymptotic charges (as defined in \cite{Prabhu19}) associated with BMS supertranslation symmetries are well-defined at $\mathcal{I}^{\pm}$ if and only if the initial data satisfy extra regularity conditions. The regularity conditions can be imposed on the free conformal initial data. Finally, given initial data that satisfy the regularity conditions, the asymptotic charge $\mathcal{Q}_{l,m}$ associated with a given harmonic $Y_{l,m}$ at $\mathcal{I}^{+}$ is related to the corresponding asymptotic charge at $\mathcal{I}^{-}$ by:  $\mathcal{Q}_{l,m}|_{\mathcal{I}^{+}}= (-1)^{l} \mathcal{Q}_{l,m}|_{\mathcal{I}^{-}}$.}
\end{center}
The structure of this paper is as follows: In Section \ref{Section: The spin-2 asymptotic charges on Minkowski spacetime}, the calculation of the spin-2 supertranslation asymptotic charges on Minkowski spacetime using Friedrich's formulation is reviewed. We start by introducing Friedrich's representation and Friedrich-gauge (F-gauge) of spatial infinity on Minkowski spacetime in Section \ref{Section: Minkowski spacetime in the F-gauge}. Since the asymptotic charges are expressed in terms of the so-called Newman-Penrose gauge (NP-gauge), Section \ref{Section:NP} provides a brief discussion of the transformation from the NP-gauge to the F-gauge. Finally, Section \ref{Section: The asymptotic charges in full GR} presents a brief description of the tools and techniques used to analyse the supertranslation asymptotic charges in full GR.

\subsection*{Notations and conventions}
This article will use tensors and spinors separately in various calculations. The following indices will be used:
\begin{itemize}
    \item $a,\,b,\,c,\ldots$: spacetime abstract tensorial indices.
    \item $i,\,j,\,k,\ldots$: spatial abstract indices
    \item $\mu,\,\nu,\dots$: spacetime coordinate indices.
    \item $\alpha,\,\beta,\dots$: spatial coordinate indices.
    \item $\mathcal{A}, \mathcal{B}, \mathcal{C}, \ldots$: coordinate indices on a 2-sphere.
    \item $A,\,B,\,C,\ldots$: abstract spinorial indices.
\end{itemize}
The components of a tensor $T_{ab}$ with respect to a tensorial frame $\{ \bme_{\bma} \}$ are defined as
\begin{equation*}
    T_{\bma \bmb} = T_{ab} \bme_{\bma}{}^{a} \bme_{\bmb}{}^{b}.
\end{equation*}
Similarly, if $\{ \bmo, \bmiota \}$ is a spin basis defined by 
\begin{equation*}
    o^{A} \equiv \bmepsilon_{\bmzero}{}^{A}, \qquad \iota^{A} \equiv \bmepsilon_{\bmone}{}^{A},
\end{equation*}
then the components of a spinor $\xi_{A}$ with respect to the spin frame $\{ \bmepsilon_{\bmA} \}$ are given by
\begin{equation*}
    \xi_{\bmA} = \xi_{A} \bmepsilon_{\bmA}{}^{A}.
\end{equation*}
The spin basis $\{ \bmo, \bmiota \}$ satisfies $\llbracket o, \iota \rrbracket =1$, where $\llbracket .,.\rrbracket$ is the antisymmetric product defined by
\begin{equation*}
\llbracket \zeta, \lambda \rrbracket = \zeta_B \lambda^B = \epsilon_{AB} \zeta^A \lambda^B.
\end{equation*}
Here, $\epsilon_{AB}$ is the antisymmetric $\epsilon$-spinor that can be regarded as a raising/lowering object for spinor indices.

\section{The spin-2 asymptotic charges on Minkowski spacetime}
\label{Section: The spin-2 asymptotic charges on Minkowski spacetime}
In this section, we summarise the calculation of supertranslation asymptotic charges for the spin-2 field as presented in \cite{MohamedKroon22}. Given the role of Friedrich's representation of spatial infinity in this calculation, let us begin by introducing the F-gauge on Minkowski spacetime.

\subsection{The Minkowski spacetime in the F-gauge}
\label{Section: Minkowski spacetime in the F-gauge}
\begin{figure}[t]
\centering
\includegraphics[width=95mm]{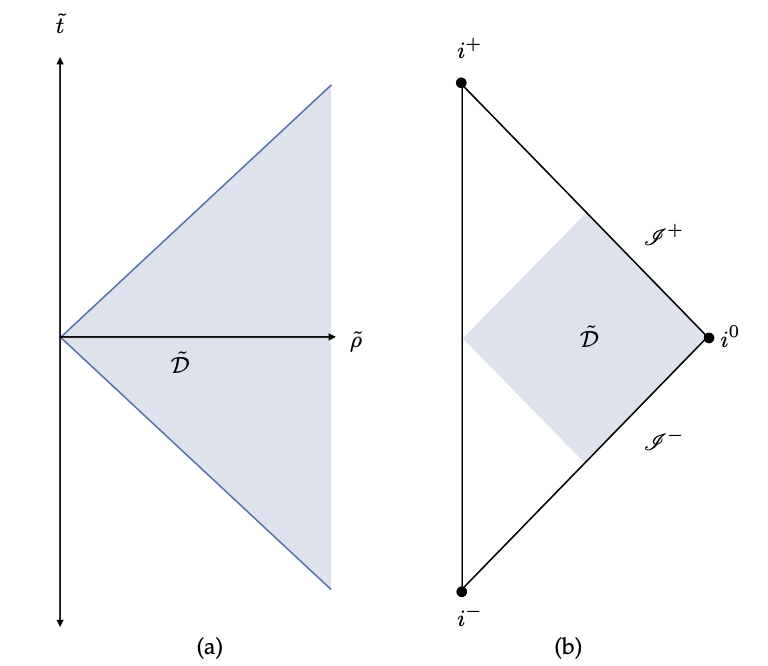}
\caption{ (a) The domain $\tilde{\mathcal{D}}$ containing spatial infinity, (b) The domain $\tilde{\mathcal{D}}$ on the conformal diagram of Minkowski spacetime.}
\label{Fig:Domain-D}
\end{figure}
To introduce the F-gauge on Minkowski spacetime $(\mathbb{R}^4,\tilde{\bmeta})$, start with the Minkowski metric $\tilde{\bmeta}$ in the standard Cartesian coordinates $(\tilde{x}^\mu)$  
\begin{equation*}
    \tilde{\bmeta} = \tilde{\eta}_{\mu\nu} \bmd{\tilde{x}^{\mu}} \otimes
   \bmd{\tilde{x}^{\nu}},
\end{equation*}
where $\tilde{\eta}_{\mu \nu} = \text{diag}(1,-1,-1,-1)$. This metric can be written in terms of the standard spherical coordinates $(\tilde{t},\tilde{\rho},\theta^{\mathcal{A}})$ as
\begin{equation*}
    \tilde{\bmeta} = \bmd{\tilde{t}}\otimes\bmd{\tilde{t}} - \bmd{\tilde{\rho}}\otimes\bmd{\tilde{\rho}}-\tilde{\rho}^2 \bmsigma,
\end{equation*}
where $\theta^{\mathcal{A}}$ is a choice of coordinates on $\mathbb{S}^{2}$ and $\bmsigma$ is the standard round metric on $\mathbb{S}^2$. Now, define $\tilde{X}^2 \equiv \tilde{\eta}_{\mu \nu} \tilde{x}^{\mu} \tilde{x}^{\nu} = \tilde{t}^2 - \tilde{\rho}^2$, where $\tilde{x}^{0} \equiv \tilde{t}$ and $\tilde{\rho}^2 \equiv (\tilde{x}^{1})^2 + (\tilde{x}^{2})^2 + (\tilde{x}^{3})^2$. Then, it is clear to see that spatial infinity is contained in the domain $\tilde{\mathcal{D}}$ (See Figure \ref{Fig:Domain-D}) defined as 
\begin{equation*}
    \tilde{\mathcal{D}} \equiv \{ p \in \mathbb{R}^4 | \hspace{1mm} \tilde{\eta}_{\mu \nu} \tilde{x}^{\mu}(p) \tilde{x}^{\nu}(p) <0 \}.
\end{equation*}
The conformal metric $\hat{\bmeta} = \Xi^{2} \tilde{\bmeta}$, with $\Xi = \tilde{X}^{-2}$ produces a point compactification of the physical spacetime $(\mathbb{R}^4, \tilde{\bmeta})$, where all the points at infinite spatial distances in the physical spacetime $(\mathbb{R}^4,\tilde{\bmeta})$ are mapped to the spatial infinity point $i^0$ in $(\mathbb{R}^4, \hat{\bmeta})$. The metric $\hat{\bmeta}$ can be written explicitly as
\begin{equation*}
    \hat{\bmeta} = \bmd{t} \otimes \bmd{t} - \bmd{\rho} \otimes \bmd{\rho} - \rho^2 \bmsigma,
\end{equation*}
where 
\begin{equation*}
    t = - \frac{\tilde{t}}{\tilde{t}^2-\tilde{\rho}^2}, \qquad \rho = - \frac{\tilde{\rho}}{\tilde{t}^2-\tilde{\rho}^2}.
\end{equation*}
To introduce Friedrich's blow-up of spatial infinity, define a new time coordinate $\tau = t/ \rho$ and the rescaling 
\begin{equation*}
    \bmeta = \frac{1}{\rho^2} \hat{\bmeta},
\end{equation*}
so that
\begin{equation}
    \bmeta = \bmd{\tau} \otimes \bmd{\tau} + \frac{\tau}{\rho} (\bmd{\tau} \otimes \bmd{\rho} + \bmd{\rho} \otimes \bmd{\tau}) - \frac{(1-\tau^2)}{\rho^2} \bmd{\rho} \otimes \bmd{\rho} - \bmsigma.
    \label{Fredrich-metric-Minkowski}
\end{equation}
From this, the relation between $\bmeta$ and $\tilde{\bmeta}$ can be written as
\begin{equation}
    \bmeta = \Theta^2 \tilde{\bmeta}, \qquad \Theta = \rho (1-\tau^2).
    \label{F-Conformal-rescaling-Minkowski}
\end{equation}
In this representation, one sees that the spacetime metric $\bmeta$ is singular at $\rho=0$ while the intrinsic metric on the $\rho=\text{const.}$ hypersurfaces have a well-defined limit as $\rho \to 0$ and is given by
\begin{equation*}
    \bmq = \bmd{\tau} \otimes \bmd{\tau} - \bmsigma.
\end{equation*}

\begin{figure}[t]
\centering
\includegraphics[width=120mm]{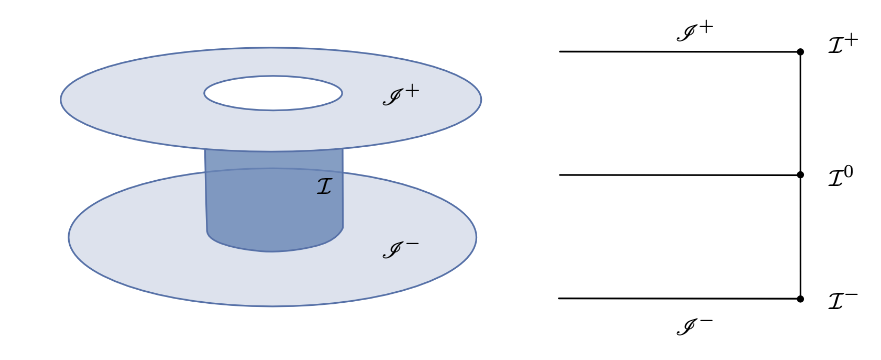}
\caption{A diagram of the neighbourhood of spatial infinity in Friedrich's representation. In this representation, the spatial infinity point $i^0$ is blown up to a cylinder $\mathcal{I}$ connecting past null infinity $\mathscr{I}^{+}$ and future null infinity $\mathscr{I}^{-}$. The critical sets $\mathcal{I}^{\pm}$ represents the sets where $\mathcal{I}$ touches $\mathscr{I}^{\pm}$. The set $\mathcal{I}^{0}$ represents the intersection of the cylinder $\mathcal{I}$ with the initial hypersurface $\{  \tau =0\}$.}
\label{Fig:Cylinder}
\end{figure}
Given the above, define the conformal extension $(\mathcal{M}, \bmeta)$ with 
\begin{equation*}
    \mathcal{M} \equiv \{ p \in \mathbb{R}^4| -1 \leq \tau(p) \leq 1, \rho(p) \geq 0 \},
\end{equation*}
then introduce the following subsets of the conformal boundary $(\Theta =0)$ --- see Figure \ref{Fig:Cylinder}.
\begin{subequations}
    \begin{align*}
        \mathscr{I}^\pm \equiv \big\{ p\in \mathcal{M} \hspace{1mm}\rvert \hspace{1mm} \tau(p) =\pm 1 \big\},  \qquad & \text{past and future null infinity} \\
        \mathcal{I} \equiv \big\{ p \in \mathcal{M} \hspace{1mm} \rvert   \hspace{1mm}  |\tau(p)|<1, \hspace{1mm} \rho(p)=0\big\}, \qquad & \text{the cylinder at spatial infinity} \\
        \mathcal{I}^{\pm} \equiv \big\{ p\in \mathcal{M} \hspace{1mm} \rvert \hspace{1mm} \tau(p)= \pm 1, \hspace{1mm} \rho(p)=0 \big\}, \qquad & \text{the critical sets at of null infinity}
    \end{align*}
\end{subequations}
and 
\begin{equation*}
    \mathcal{I}^{0} \equiv \big\{ p \in \mathcal{M}\hspace{1mm} \rvert \hspace{1mm} \tau(p)=0, \hspace{1mm} \rho(p)=0\big\},
\end{equation*}
where $\mathcal{I}^0$
is the intersection of $\mathcal{I}$ with the initial hypersurface
$\mathcal{S}_{*} \equiv \{ \tau = 0 \}$. In subsequent discussions, we will refer to $(\mathcal{I},\bmq)$ as the cylinder at spatial infinity. Moreover, it will be convenient to introduce a frame basis $\{ \bme_{\bma} \}$ adapted to Friedrich's cylinder at spatial infinity on Minkowski spacetime, the so-called F-gauge frame. Start with the Minkowski metric $\bmeta$ given by \eqref{Fredrich-metric-Minkowski}. It is straightforward to see that the metric on the hypersurfaces $\mathcal{Q}_{\tau,\varrho}$ of constant $\rho$ and $\tau$ is the standard metric on $\mathbb{S}^{2}$. Then, introduce the complex null frame $\{ \bmpartial_{+}, \bmpartial_{-} \}$ on $\mathcal{Q}_{\tau,\rho}$ and propagate $\{ \bmpartial_{+}, \bmpartial_{-} \}$ off $\mathcal{Q}_{\tau,\rho}$ by imposing 
\begin{equation*}
    [ \bmpartial_{\tau}, \bmpartial_{\pm} ] =0, \qquad [ \bmpartial_{\rho}, \bmpartial_{\pm} ]=0.
\end{equation*}
Now, the F-gauge frame $\{ \bme_{\bmA \bmA'} \}$ and their dual $\{ \bmomega^{\bmA \bmA'} \}$ can be defined as follows
\begin{subequations}
    \begin{equation*}
        \bme_{\bmzero \bmzero'} = \frac{\sqrt{2}}{2} \left( (1-\tau) \bmpartial_{\tau} + \rho \bmpartial_{\rho} \right), \qquad \bmomega^{\bmzero \bmzero'} = \frac{\sqrt{2}}{2} \left( \bmd{\tau} - \frac{1}{\rho} (1-\tau) \bmd{\rho} \right), 
    \end{equation*}
    \begin{equation*}
        \bme_{\bmone \bmone'} = \frac{\sqrt{2}}{2} \left( (1+\tau) \bmpartial_{\tau} - \rho \bmpartial_{\rho} \right), \qquad \bmomega^{\bmone \bmone'} = \frac{\sqrt{2}}{2} \left( \bmd{\tau} + \frac{1}{\rho} (1+\tau) \bmd{\rho} \right), 
    \end{equation*}
    \begin{equation*}
        \bme_{\bmzero \bmone'} = \frac{\sqrt{2}}{2} \bmpartial_+, \qquad \bmomega^{\bmzero \bmone'} = \sqrt{2} \bmomega^+,
    \end{equation*}
    \begin{equation*}
        \bme_{\bmone \bmzero'} = \frac{\sqrt{2}}{2} \bmpartial_-, \qquad \bmomega^{\bmone \bmzero'} = \sqrt{2} \bmomega^-,
    \end{equation*}
\end{subequations}
where $\bme_{\bmA \bmA'}$ is obtained from $\bme_{\bma}$ by contraction with the Infeld-van der Waerden symbols $\sigma^{\bma}{}_{\bmA \bmA'}$. So, $ \bme_{\bmA \bmA'} \equiv \sigma^{\bma}{}_{\bmA \bmA'} \bme_{\bma}$. The dual frame $\bmomega^\pm$ satisfy
\begin{equation*}
    \langle \bmomega^+,\bmpartial_+\rangle =1, \qquad \langle \bmomega^-,\bmpartial_-\rangle =1.
\end{equation*}
In terms of the above frame fields, the metric $\bmeta$ can be written as
$$
\bmeta = \epsilon_{\bmA \bmB} \epsilon_{\bmA' \bmB'} \bmomega^{\bmA \bmA'} \otimes \bmomega^{\bmB \bmB'}.
$$
\begin{remark}
    A special class of conformal curves, known as conformal geodesics, play a major role in Friedrich's representation of spatial infinity --- see \cite{MohamedKroon21} for details. For example, on Minkowski spacetime, the curves of constant $\rho$ and $\theta^{\mathcal{A}}$ describe a non-intersecting congruence of conformal geodesics on $\mathcal{M}$, suggesting that the congruence of conformal geodesics starting at $\rho=0$ coincide with the cylinder at spatial infinity. One of the remarkable properties of conformal geodesics is that they specify a canonical conformal factor $\Theta$ \cite{kroon16, Friedrich95}, and an F-gauge metric $\bmg$ given by
    \begin{equation}
        \bmg = \Theta^2 \tilde{\bmg},
    \end{equation}
    where $\tilde{\bmg}$ is the metric on a vacuum Einstein spacetime $(\tilde{\mathcal{M}},\tilde{\bmg})$. On Minkowski spacetime, the canonical conformal factor associated with the above-mentioned congruence of conformal geodesics on Minkowski is equivalent to $\Theta$ given in \eqref{F-Conformal-rescaling-Minkowski}.
\end{remark}
Further discussion of Friedrich's blow-up of spatial infinity and the F-gauge on more general spacetimes will be postponed for later sections.

\subsection{NP-gauge to F-gauge}
\label{Section:NP}
As mentioned in the introduction, the supertranslation asymptotic charges at $\mathscr{I}^{\pm}$ are generally expressed in terms of the NP-gauge, comprised of certain conformal gauge conditions, certain coordinates and an orthonormal frame field $\{ \bme^{\bullet}_{\bma} \}$ satisfying certain frame gauge conditions. A general description of these gauge conditions was given in \cite{FriedrichKannar00}, along with a prescription of the transformation between the NP-gauge and the F-gauge. The main observation in \cite{FriedrichKannar00} is that the NP-gauge frame is adapted to null infinity $\mathscr{I}^{\pm}$ while the F-gauge frame is adapted to Cauchy hypersurfaces. Additionally, the NP conformal gauge conditions, as described in \cite{FriedrichKannar00}, imply that a metric $\bmg^{\bullet}$ satisfying those conditions will be related to the F-gauge metric $\bmg$ by
\begin{equation*}
    \bmg^{\bullet} = \theta^2 \bmg,
\end{equation*}
with $\theta$ satisfying a linear ordinary differential equation, which can be solved on the generators of $\mathscr{I}^{\pm}$ --- see \cite{FriedrichKannar00}. Given the above, the NP-gauge orthonormal frame $\{ \bme^{\bullet}_{\bmA \bmA'} \}$, defined by $\bme^{\bullet}_{\bmA \bmA'} \equiv \sigma^{\bma}{}_{\bmA \bmA'} \bme^{\bullet}_{\bma}$, is related to the F-gauge orthonormal frame $\{ \bme_{\bmA \bmA'} \}$ by
\begin{equation*}
    \bme^{\bullet}_{\bmA \bmA'} = \theta^{-1} \Lambda^{\bmB}{}_{\bmA} \bar{\Lambda}^{\bmB'}{}_{\bmA'} \bme_{\bmB \bmB'}, \qquad \Lambda^{\bmB}{}_{\bmA} \in \text{SL}(2,\mathbb{C}). 
\end{equation*}
The general framework in \cite{FriedrichKannar00} was used in \cite{GasperinKroon20} to obtain an explicit transformation between the NP-gauge and the F-gauge on Minkowski spacetime. In particular, the results in \cite{GasperinKroon20} show that $\theta=1$ and 
\begin{subequations}
    \begin{eqnarray*}
        && \Lambda^{\bmone}{}_{\bmzero} = \frac{2 e^{i \omega}}{\sqrt{\rho}(1+\tau)}, \qquad \Lambda^{\bmzero}{}_{\bmone} = \frac{e^{-i \omega} \sqrt{\rho} (1+\tau)}{2}, \qquad \Lambda^{\bmone}{}_{\bmone} = \Lambda^{\bmzero}{}_{\bmzero} =0, \qquad \text{on } \mathscr{I}^{+} \\
        && \Lambda^{\bmone}{}_{\bmzero} = \frac{e^{-i \omega} \sqrt{\rho} (1-\tau)}{2} , \qquad \Lambda^{\bmzero}{}_{\bmone} = \frac{2 e^{i \omega}}{\sqrt{\rho}(1-\tau)}, \qquad \Lambda^{\bmone}{}_{\bmone} = \Lambda^{\bmzero}{}_{\bmzero} =0, \qquad \text{on } \mathscr{I}^{-}
    \end{eqnarray*}
\end{subequations}
where $\omega$ is an arbitrary real number that encodes the spin rotation of the frames on $\mathbb{S}^2$. Given the above transformation between the NP-gauge and the F-gauge on Minkowski spacetime, the asymptotic charges can be evaluated on the critical sets $\mathcal{I}^{\pm}$ given solutions for the spin-2 field equations. 

\subsection{The spin-2 charges in the F-gauge}
The goal of this section is to obtain an expression of the supertranslation asymptotic charges that can be evaluated at the critical sets $\mathcal{I}^{\pm}$. First, introduce the NP null tetrad $(l^{a},n^{a},m^{a}, \bar{m}^{a})$ as
\begin{equation}
        l^{a} \equiv \bme^{\bullet}_{\bmzero \bmzero'}, \qquad n^{a} \equiv \bme^{\bullet}_{\bmone \bmone'}, \qquad m^{a} \equiv \bme^{\bullet}_{\bmzero \bmone'}, \qquad \bar{m}^{a} \equiv \bme^{\bullet}_{\bmone \bmzero'}. 
        \label{NP-tetrad-in-spinors}
    \end{equation}
Then, let $W^{\bullet}_{abcd}$ denote a Weyl-like tensor, i.e., a tensor with symmetries of the Weyl tensor, and define $\mathcal{W}^{\bullet}_{abcd}$ as
\begin{equation*}
    \mathcal{W}^{\bullet}_{abcd} \equiv W^{\bullet}_{abcd} + i ({}^{*}W)^{\bullet}_{abcd},
\end{equation*}
where $({}^{*}W)^{\bullet}_{abcd}$ is the left Hodge dual of $W^{\bullet}_{abcd}$. Then, the spinorial counterpart of $W^{\bullet}_{abcd}$ can be decomposed in terms of the symmetric spin-2 spinor $\psi^{\bullet}_{ABCD}$ as 
\begin{equation}
    W^{\bullet}_{AA'BB'CC'DD'} = -\psi^{\bullet}_{ABCD} \epsilon^{\bullet}_{A'B'} \epsilon^{\bullet}_{C'D'} - \Bar{\psi}^{\bullet}_{A'B'C'D'} \epsilon^{\bullet}_{AB} \epsilon^{\bullet}_{CD}.
    \label{Weyl-like-tensor-in-spinors}
\end{equation}
\begin{remark}
    The ${}^{\bullet}$ notation indicates that $W^{\bullet}_{abcd}$ (or $\mathcal{W}^{\bullet}_{abcd}$), $\epsilon^{\bullet}_{AB}$ are in the NP-gauge. Given that $\theta=1$ on Minkowski spacetime, one has $W^{\bullet}_{abcd} = W_{abcd}$. However, this is not true for general spacetimes, i.e., the Weyl tensor $C^{\bullet}_{abcd}$ associated with $\bmg^{\bullet}$ will be related to $C_{abcd}$ associated with $\bmg$ by $C^{\bullet}_{abcd} = \theta^2 C_{abcd}$. Moreover, the $\epsilon$-spinor in the NP-gauge $\epsilon^{\bullet}_{AB}$ will be related to $\epsilon_{AB}$ by $\epsilon^{\bullet}_{AB} = \theta \epsilon_{AB}$. If one wishes to obtain a relation between $C^{\bullet}_{\bma \bmb \bmc \bmd}$ and $C_{\bma \bmb \bmc \bmd}$, where $C^{\bullet}_{\bma \bmb \bmc \bmd} \equiv C^{\bullet}_{abcd} \bme^{\bullet}_{\bma}{}^{a} \bme^{\bullet}_{\bmb}{}^{b} \bme^{\bullet}_{\bmc}{}^{c} \bme^{\bullet}_{\bmd}{}^{d}$ and $C_{\bma \bmb \bmc \bmd} \equiv C_{abcd} \bme_{\bma}{}^{a} \bme_{\bmb}{}^{b} \bme_{\bmc}{}^{c} \bme_{\bmd}{}^{d}$, then one makes use of the transformation between $C^{\bullet}_{abcd}$ and $C_{abcd}$ as well as the transformation between the NP-gauge frame $\{ \bme^{\bullet}_{\bma} \}$ and the F-gauge frame $\{ \bme_{\bma} \}$ given by $\bme^{\bullet}_{\bma}= \theta^{-1} \Lambda^{\bmb}{}_{\bma} \bme_{\bmb}$, where $\Lambda^{\bmb}{}_{\bma} \in \text{O}(1,3)$.
\end{remark}
Following the discussion in \cite{Prabhu19}, the asymptotic charges associated with smooth functions $\lambda$ on $\mathbb{S}^2$ can be written as 
\begin{equation*}
    \mathscr{Q} = \int_{\mathcal{C}} \lambda \mathcal{W}^{\bullet}_{abcd} l^{a} n^{b} m^{c} \bar{m}^d \bmd{S},
\end{equation*}
where $\mathcal{C}$ denotes a cross-section of $\mathscr{I}^{\pm}$. From \eqref{NP-tetrad-in-spinors} and \eqref{Weyl-like-tensor-in-spinors}, it can be shown that the charges $\mathscr{Q}$ can be written as 
\begin{equation*}
    \mathscr{Q} = - 2 \int_{\mathcal{C}} \lambda \bar{\psi}^{\bullet}_{2} \bmd{S},
\end{equation*}
where $\bar{\psi}^{\bullet}_{2} \equiv \bar{\psi}^{\bullet}_{\bmzero' \bmzero' \bmone' \bmone'}$. 
To evaluate the charges at the critical sets, one must obtain an expression for $\mathscr{Q}$ in terms of the F-gauge. The transformation from the NP-gauge to the F-gauge on Minkowski spacetime, discussed in the previous section, implies that
\begin{equation*}
    \psi^{\bullet}_{2} = \psi_{2},
\end{equation*}
where $\psi_{2} \equiv \psi_{\bmzero \bmzero \bmone \bmone}$. Thus, the final expression of the charges in the F-gauge is given by
\begin{equation}
    \mathscr{Q} = - 2 \int_{\mathcal{C}} \lambda \bar{\psi}_{2} \bmd{S}.
    \label{Final-expression-spin-2-charges}
\end{equation}
To evaluate this expression at $\mathcal{I}^{\pm}$, the next step is to obtain a solution for $\bar{\psi}_{2}$ using the field equations. 
\subsection{The spin-2 field equations}
The spinorial spin-2 field equation can be written as $\nabla^A{}_{A'} \psi_{ABCD} = 0$. Applying $-2 \nabla_{E}{}^{A'}$, the wave equation satisfied by $\psi_{ABCD}$ can be written as
\begin{equation}
    \square \psi_{ABCD} =0,
    \label{Spin-2-field-equation}
\end{equation}
where $\square \equiv \nabla_{AA'}\nabla^{AA'}$ is the D'Alembertian operator. To analyse the solutions for this equation in a neighbourhood of spatial infinity, assume that the components $\psi_{n}$ of the spin-2 spinor can be expanded near $\rho=0$ in terms of spin-weighted spherical harmonics ${}_{n}Y_{l,m}$ as 
\begin{equation}
    \psi_{n} = \sum_{l=|2-n|}^{\infty} \sum_{m=-l}^{l} a_{n;l,m}(\tau) {}_{2-n}Y_{l,m} + o_1(\rho), \qquad \text{for } n =0, \ldots,4.
    \label{expansion-spin-2-field}
\end{equation}
where $a_{n;l,m}: \mathbb{R} \rightarrow \mathbb{C}$. 
\begin{remark}
The expansion \eqref{expansion-spin-2-field} is consistent with the estimates developed in \cite{Friedrich03-2} that demonstrates that for a certain non-trivial class of initial data, the components $\psi_{n}$ can be expanded as
    \begin{equation*}
        \psi_n = \sum_{k=|2-n|}^{p-1} \frac{1}{k!}\psi_n^{(k)}\rho^k + R_p[\psi_n].
    \end{equation*}
  In the above, the coefficients $\psi_n^{(k)}$ are explicitly known functions of $\tau$ and the angular variables which are smooth for $\tau\in(-1,1)$ and whose regularity at $\tau=\pm 1$ can be controlled in terms of the initial data. The reminder satisfies $R_p[\psi_n]\in C^m$ for $p\geq m+6$ for $\rho$ near 0 and $\tau\in [-1,1]$. 
\end{remark}
Using \eqref{expansion-spin-2-field} and substituting in \eqref{Spin-2-field-equation}, one obtains second order ordinary differential equations for the coefficients $a_{n;l,m}(\tau)$
\begin{subequations}
\begin{equation}
    (1-\tau^2) \ddot{a}_{0;l,m} + 2 (2-\tau) \dot{a}_{0;l,m} + l (l+1) a_{0;l,m} =0, \label{WaveEqnSpin2Expansion1} \\
\end{equation}
\begin{equation}
    (1-\tau^2) \ddot{a}_{1;l,m} + 2(1-\tau) \dot{a}_{1;l,m} +l (l+1) a_{1;l,m} =0, \label{WaveEqnSpin2Expansion2} \\
\end{equation}
\begin{equation}
    (1-\tau^2) \ddot{a}_{2;l,m} - 2\tau \dot{a}_{2;l,m} + l (l+1) a_{2;l,m} =0, \label{WaveEqnSpin2Expansion3}  \\
\end{equation}
\begin{equation}
    (1-\tau^2) \ddot{a}_{3;l,m} -2(1+\tau) \dot{a}_{3;l,m} + l (l+1) a_{3;l,m}=0, \label{WaveEqnSpin2Expansion4} \\
\end{equation}
\begin{equation}
    (1-\tau^2) \ddot{a}_{4;l,m} - 2(2+\tau) \dot{a}_{4;l,m} + l (l+1) a_{4;l,m}=0. \label{WaveEqnSpin2Expansion5}
\end{equation}
\end{subequations}
Now, assume that the initial data $(\psi_{n})|_{\mathcal{S}_{*}}$ on $\mathcal{S}_{*} \equiv \{ \tau=0 \}$ can be expanded near $\rho=0$ as
\begin{equation*}
    \psi_n|_{\mathcal{S}_{*}} = \sum_{l=|2-n|}^{\infty} \sum_{m=-l}^{l} a_{n;l,m}(0) {}_{2-n} Y_{l,m} + o(\rho).
    \label{spin2-initial-data}
\end{equation*} 
Given that the expression of the charges \eqref{Final-expression-spin-2-charges} is written in terms of $\psi_{2}$, one only requires the solution for \eqref{WaveEqnSpin2Expansion3} in order to evaluate $\mathscr{Q}$ at $\tau = \pm 1$. Equation \eqref{WaveEqnSpin2Expansion3} is a Jacobi differential equation, with a solution that can be expressed in terms of hypergeometric functions --- see \cite{Weisstein}. However, a simpler expression can be obtained by using the differential equation solver of the Wolfram Language. In particular, it can be shown that for $ l \geq 0$ and $-l \leq m \leq l$, the solution $a_{2;l,m}$ is given by
\begin{equation}
    a_{2;l,m}(\tau) = A_{l,m} P_{l}(\tau) + B_{l,m} Q_{l}(\tau),
    \label{Solution-a2}
\end{equation}
where $P_{l}(\tau)$ is the Legendre polynomial of order $l$ and
$Q_{l}(\tau)$ is the Legendre function of the second kind of order
$l$. The constants $A_{l,m}$ and $B_{l,m}$ can be expressed in terms of the initial data for $a_{2;l,m}$. Since $Q_{l}(\tau)$ is proportional to $\ln{(1+\tau)}$ and $\ln{(1-\tau)}$, the solution \eqref{Solution-a2} will diverge logarithmically near $\tau = \pm 1$, unless $B_{l,m}=0$.

To obtain a well-defined solution for $a_{2;l,m}$ at the critical sets, the constant $B_{l,m}$ is required to vanish. Note that $B_{l,m}$ can be written in terms of $a_{2;l,m}(0)$ and $\dot{a}_{2;l,m}(0)$ as
\begin{equation}
        B_{l,m} = \frac{\sqrt{\pi} (l+1)}{\Gamma(-\frac{l}{2}) \Gamma(\frac{l+3}{2})} a_{2;l,m}(0) + \frac{\sqrt{\pi}}{\Gamma(\frac{1}{2}-\frac{l}{2}) \Gamma(1+\frac{l}{2})} \dot{a}_{2;l,m}(0), \label{Spin2-Blm}
\end{equation}
where $\Gamma$ denotes the Gamma function. Then, the observation that the coefficient of $a_{2;l,m}(0)$ vanishes for even $l$ while the coefficient of $\dot{a}_{2;l,m}(0)$ vanishes for odd l suggests the following regularity conditions:
\begin{lemma}
The solution \eqref{Solution-a2} is well-defined at $\mathcal{I}^{\pm}$ if and only if: \begin{enumerate}
    \item $a_{2;l,m}(0)=0$ for odd $l$, and
    \item $\dot{a}_{2;l,m}(0) =0$ for even $l$.
\end{enumerate}
\label{Initial-data-Spin2}
\end{lemma}
These regularity conditions can be expressed in terms of freely specifiable data as shown in \cite{MohamedKroon22}. 

Making use of \eqref{Final-expression-spin-2-charges}, \eqref{expansion-spin-2-field} and \eqref{Solution-a2} and by choosing initial data satisfying Lemma \ref{Initial-data-Spin2} and $\lambda = Y_{l,m}$, the charge $\mathscr{Q}_{l,m}$ associated with $Y_{l,m}$ at $\mathcal{I}^{\pm}$ can be written as
\begin{equation}
\mathscr{Q}_{l,m}|_{\mathcal{I}^{\pm}} = \begin{cases} 
         2 (l+1) Q_{l+1}(0) (a_2)_* \qquad \text{for even }l \geq 0, \\
         \pm \sqrt{l(l+1)} Q_{l}(0) \left( (a_1)_* - (a_3)_* \right) \qquad \text{for odd }l,
    \end{cases}
\end{equation}
where $(a_n)_{*} \equiv a_{n;l,m}(0)$. The main conclusions from the above discussion are
\begin{enumerate}
    \item For generic boosted initial data, the charges $\mathscr{Q}$ are not well-defined in the limits of spatial infinity, i.e., at the critical sets $\mathcal{I}^{\pm}$.
    \item Boosted initial data satisfying Lemma \ref{Initial-data-Spin2} allows us to obtain well-defined expressions for $\mathscr{Q}$ at the critical sets. 
    \item The antipodal matching of the charges is obtained naturally in this formalism. In particular, we have $\mathscr{Q}_{l,m}|_{\mathcal{I}^{+}} = (-1)^{l} \mathscr{Q}_{l,m}|_{\mathcal{I}^{-}}$.
\end{enumerate}

\section{The asymptotic charges in full GR}
\label{Section: The asymptotic charges in full GR}
The process of the calculation of the asymptotic charges for the spin-2 field at the critical sets presented in the previous section can be extended to the full GR setting. For this, assume that $(\tilde{\mathcal{M}},\tilde{\bmg})$ is a spacetime satisfying the vacuum Einstein field equations i.e.
\begin{equation}
    \tilde{R}_{ab} =0,
    \label{Vacuum-Einstein-field-eqs}
\end{equation}
where $\tilde{R}_{ab}$ is the Ricci tensor associated with the Levi-Civita connection $\tilde{\bmnabla}$ of $\tilde{\bmg}$. The conformal rescaling 
\begin{equation}
    \bmg = \Xi^{2} \tilde{\bmg},
    \label{conformal-rescaling}
\end{equation}
implies transformation laws for the physical fields e.g. the curvature tensor $\tilde{R}^{a}{}_{bcd}$, the Schouten tensor $\tilde{L}_{ab}$ etc. It follows that the vacuum Einstein field equations are not conformally invariant and that the field equations implied by \eqref{conformal-rescaling} cannot be analysed at the conformal boundary $\Xi=0$ since the conformal Ricci tensor $R_{ab}$ is singular at the points where $\Xi=0$. If $\tilde{C}^{a}{}_{bcd}$ denotes the Weyl tensor, then the Bianchi identity can be written in terms of the Levi-Civita connection associated with $\bmg$ as
\begin{equation}
    \nabla_{a} ( \Xi^{-1} \tilde{C}^{a}{}_{bcd}) =0.
    \label{Bianchi-equation}
\end{equation}
If one defines the rescaled Weyl tensor $d^{a}{}_{bcd} \equiv \Xi^{-1} \tilde{C}^{a}{}_{bcd}$, then equation \eqref{Bianchi-equation} can be written as 
\begin{equation}
    \nabla_{a} d^{a}{}_{bcd} =0.
\end{equation} 
Exploiting the symmetries of the rescaled Weyl tensor implies
\begin{equation}
    \nabla_{[e} d^{a}{}_{|b|cd]} =0.
\end{equation}
Our calculations of the asymptotic charges rely on Friedrich's extended conformal field equations \cite{Friedrich81a,Friedrich81b,Friedrich83,kroon16} written in terms of a Weyl connection $\hat{\bmnabla}$ satisfying
\begin{equation*}
    \hat{\nabla}_{a} g_{bc} = -2 f_{a} g_{bc},
\end{equation*}
where $f_{a}$ is an arbitrary 1-form. The explicit form of these equations will not be necessary for this article, interested readers can refer to Chapter 8 in \cite{kroon16}. The extended conformal field equations yield differential equations to be solved for the $\bmg$-orthonormal frame fields $\{ \bme_{\bma} \}$, the components of the Weyl connection coefficients $\hat{\Gamma}_{\bma}{}^{\bmb}{}_{\bmc}$, the Schouten tensor $\hat{L}_{\bma \bmb}$ and the rescaled Weyl tensor $d^{\bma}{}_{\bmb \bmc \bmd}$. One significant feature of the extended conformal field equations is that they exhibit gauge freedom indicated by the fact that there are no equations to fix the conformal factor $\Xi$ and the Weyl connection $\hat{\bmnabla}$. To fix this gauge freedom, one can make use of the so-called conformal Gaussian gauge, based on conformal geodesics, that allows us to write the field equations as a symmetric hyperbolic system in which the evolution equations reduce to a transport system along the conformal geodesics. Given the field equations in this gauge, it is possible to obtain a spinorial version of these equations to be analysed near spatial infinity. 

The above discussion highlights one of the key tools of conformal methods in GR. The following section will introduce Friedrich's regular initial value problem for the conformal field equations.
\subsection{Friedrich's regular initial value problem}
\label{Section:Friedrich's regular initial value problem}
The purpose of this section is to briefly introduce Friedrich's formulation in full GR. As mentioned in the introduction, the aim of Friedrich's formulation is to introduce a regular initial value problem for the conformal field equations near spatial infinity. An extensive discussion of this framework is provided in \cite{Friedrich98, FriedrichKannar00}. In this framework, the spacetime $(\tilde{\mathcal{M}},\tilde{\bmg})$ is assumed to be the development of some asymptotically Euclidean and regular \cite{Geroch72,kroon16} initial data $(\tilde{\mathcal{S}},\tilde{\bmh}, \tilde{\bmK})$. In particular, the initial data $(\tilde{\mathcal{S}},\tilde{\bmh}, \tilde{\bmK})$ is said to be an asymptotically Euclidean and regular manifold if there exists a 3-dimensional smooth compact manifold $(\mathcal{S}',\bmh')$ with a point $i \in \mathcal{S}'$, a diffeomorphism $\Phi$ from $\mathcal{S}'\setminus\{ i \}$ onto $\tilde{\mathcal{S}}$ and a conformal factor $\Omega'$ which is analytic on $\mathcal{S}'$ and satisfying i) $\Omega'=0, \bmd{\Omega'}=0$ and $\text{Hess}(\Omega')=-2 \bmh'$ at $i$, ii) $\Omega'>0$ on $\mathcal{S}' \setminus \{ i\}$, iii) $\bmh' = \Omega'^2 \Phi_{*} \tilde{\bmh}$ on $\mathcal{S}' \setminus \{ i\}$. To apply this, start with the initial data satisfying the Hamiltonian and momentum constraints as introduced in \cite{Huang10}:
\begin{proposition}
    For any $\xi, \zeta \in C^{2}(\mathbb{S}^2)$, there exists a vacuum initial data set $(\tilde{\bmh}, \tilde{\bm{K}})$ such that the components of $\tilde{\bmh}$ and $\tilde{\bm{K}}$ with respect to the standard Euclidean coordinate chart $(x^{\alpha})$ have the following asymptotics:
    \begin{subequations}
        \begin{equation}
            \tilde{h}_{\alpha \beta} = -\delta_{\alpha \beta} - \frac{1}{r} \left[ \left( A - \frac{\xi}{2} \right) \delta_{\alpha \beta} + \xi \frac{x_{\alpha} x_{\beta}}{r^2} \right] + O_2 (r^{-2}),
        \end{equation}
        \begin{equation}
            \tilde{K}_{\alpha \beta} = \frac{1}{r^2} \left[  - \frac{1}{2} \zeta \delta_{\alpha \beta} + \frac{1}{r} \left( - B_{\alpha} x_{\beta} - B_{\beta} x_{\alpha} + (B^{\gamma} x_{\gamma}) \delta_{\alpha \beta}\right) + \zeta \frac{x_{\alpha} x_{\beta}}{r^2} \right] + O_1 (r^{-3})
        \end{equation}
        \label{vacuum-initial-data}
    \end{subequations}
    where $A$, $\{ B_{\alpha}\}_{\alpha=1}^{3}$ are some constants and $r= \sqrt{(x^1)^2 + (x^2)^2 + (x^3)^2}$.
\end{proposition}
Then, define the inverse coordinates $(y^{\alpha})$ and the conformal factor $\Omega'$ as
\begin{equation*}
    y^{\alpha} = - \frac{x^{\alpha}}{r^2}, \qquad \Omega' = \frac{\varrho^2}{\sqrt{1 + A \varrho}},
\end{equation*}
so that the components of the conformal initial data $\bmh' = \Omega'^2 \tilde{\bmh}$ and $\bmK' = \Omega' \tilde{\bmK}$ can be expanded around $\varrho = \sqrt{(y^{1})^2 + (y^{2})^2+(y^{3})^2}= 0$ as
\begin{subequations}
    \begin{equation}
        h'_{\alpha \beta} = -\delta_{\alpha \beta} - \xi \varrho \left( \frac{y_{\alpha} y_{\beta}}{\varrho^2} -\frac{1}{2} \delta_{\alpha \beta} \right) + O_2 (\varrho^2), 
        \label{Initial-data-h'}
    \end{equation}
    \begin{equation}
        K'_{\alpha \beta} = - \frac{\zeta}{2} \delta_{\alpha \beta} - \frac{1}{\varrho} \left( B_{\alpha} y_{\beta} + B_{\beta} y_{\alpha}+ \frac{1}{2} (B^{\gamma}y_{\gamma}) \delta_{\alpha \beta} \right) + \left( \zeta - 4 \frac{(B^{\gamma}y_{\gamma})}{\varrho}  \right) \frac{y_{\alpha} y_{\beta}}{\varrho^2} + O_1(\varrho),
        \label{Initial-data-K'}
    \end{equation}
    \label{Initial-data-h'andK'}
\end{subequations}
The $O(\varrho)$ term in \eqref{Initial-data-h'} can be made to vanish by performing a coordinate transformation from $(y^{\alpha})$ to normal coordinates $(z^{\alpha})$ \cite{LBrewin}. Then, the term $O(\varrho^2)$ can be removed by performing a further conformal transformation
\begin{equation*}
    \Omega' \to \Omega \equiv \varpi \Omega',
\end{equation*}
where 
\begin{equation*}
    \varpi \equiv e^f, \quad \text{ with } f= \frac{1}{2} l'_{\alpha \beta}(i) z^{\alpha} z^{\beta}.
    \label{Conforaml-normal-conformal-factor}
\end{equation*}
Here, $l'_{\alpha \beta}(i)$ denotes the components of the Schouten tensor associated with $\bmh'$ in normal coordinates $(z^{\alpha})$ evaluated at $i (\varrho =0)$. If $h'^{(0)}_{\alpha \beta}$ is the metric at $i$ and $|z|^{2} \equiv h'^{(0)}_{\alpha \beta} z^{\alpha} z^{\beta}$, then the components of the conformal initial data $\bar{\bmh} = \varpi^2 \bmh'$ and $\bar{\bmK} = \varpi \bmK'$ can be written as 
\begin{subequations}
    \begin{equation*}
        \bar{h}_{\alpha \beta} = - \delta_{\alpha \beta} + O(|z|^3).
        \label{Conformal-normal-metric}
    \end{equation*}
    \begin{equation*}
        \bar{K}_{\alpha \beta} = - \frac{\zeta}{2} \delta_{\alpha \beta} - \frac{1}{2} \left( B_{\alpha} \vartheta_{\beta} + B_{\beta} \vartheta_{\alpha} + \frac{1}{2} (B^{\gamma} \vartheta_{\gamma}) \delta_{\alpha \beta} \right) + \zeta \vartheta_{\alpha} \vartheta_{\beta} - 4 (B^{\gamma} \vartheta_{\gamma}) \vartheta_{\alpha} \vartheta_{\beta} + O(|z|).
    \end{equation*}
\end{subequations}
where $\vartheta^{\alpha} = z^{\alpha}/|z|$. The initial data $(\bar{\bmh}, \bar{\bmK})$ will be referred to as the conformal normal initial data. It can be shown, using the conformal constraint equations, that the initial data for the components of the conformal Schouten tensor $\bar{L}_{\alpha \beta}$ and the electric and magnetic parts of the Weyl tensor, $\bar{d}_{\alpha \beta}$ and $\bar{d}_{\alpha \beta \gamma}$, respectively, are singular at $|z|=0$. To introduce regular initial data, one must introduce a further conformal rescaling as suggested in \cite{Friedrich98} 
\begin{equation}
    \Omega \to \kappa^{-1} \Omega,
    \label{Friedrich-conformal-rescaling}
\end{equation}
with $\kappa = O(|z|)$. Let $\rho = |z|$, then the conformal factor $\Omega$ can be expanded around $\rho=0$ as
\begin{equation*}
    \Omega = \rho^2 + \frac{1}{6} \Pi_{3}[\Omega] \rho^3 + O(\rho^4),
\end{equation*}
where $\Pi_{3}[\Omega]$ is written in terms of the angular coordinates $\vartheta^{\alpha}$, the constant $A$, the function $\alpha$ and its derivatives with respect to $\vartheta^{\alpha}$. 

The conformal rescaling \eqref{Friedrich-conformal-rescaling} introduces the conformal metric $\bmh = \kappa^{-2} \bar{\bmh}$. Then, if $\{ \bme_{\bmi} \}$ is an $\bmh$-orthonormal frame, one can show
\begin{equation*}
    h_{\bmi \bmj} = - \delta_{\bmi \bmj} + O(|z|^3), 
\end{equation*}
and
\begin{equation*}
    K_{\bmi \bmj} = O(|z|), \qquad L_{\bmi \bmj} = O(|z|), \qquad d_{\bmi \bmj} = O(1), \qquad d_{\bmi \bmj \bmk} = O(1).
\end{equation*}
Hence, the initial data $(\bmh, \bmK)$ for the conformal field equations are regular at $|z|=0$.

One of the advantages of using the conformal Gaussian gauge mentioned in the last section is that it implies a conformal factor $\Theta$ that can be written in terms of initial data. Following \cite{Friedrich95,Friedrich98}, we have
\begin{equation}
    \Theta = \kappa^{-1} \Omega \left( 1 - \tau^{2} \frac{\kappa^2}{\omega} \right),
    \label{Conformal-factor-Theta}
\end{equation}
where $\tau$ refers to the parameter along the conformal geodesics used to construct the conformal Gaussian system and
\begin{equation*}
    \omega = \frac{2 \Omega}{\sqrt{|\bmh(\bmd{\Omega},\bmd{\Omega})|}}.
\end{equation*}

\begin{remark}
    In the following, {\em $\text{SU}(2,\mathbb{C})$} refers to the special unitary group of degree $2$ over complex numbers. We also use {\em $\text{SU}(\mathcal{S})$} to refer to the bundle of normalised spin frames over a manifold $\mathcal{S}$ with structure group {\em $\text{SU}(2,\mathbb{C})$}. 
\end{remark}

The basic idea of the blow-up of the point $i$ involves replacing $i$ with the space of directions pointing out of $i$. In other words, the blow-up of $i$ is a certain subspace of the tangent space at $i$, which is diffeomorphic to $\mathbb{S}^{2}$. In Friedrich's formulation, rather than working with tensor frames, the blow-up of $i$ is achieved by considering a certain subset of the bundle of the normalised spin frames $\text{SU}(\mathcal{S}')$ with structure group $\text{SU}(2,\mathbb{C})$ --- see \cite{FriedrichKannar00} for details. In this picture, the blow-up of $i$ is diffeomorphic to $\mathbb{S}^{3}$ while its quotient by $\text{U}(1)$ is diffeomorphic to $\mathbb{S}^{2}$. The extra dimension in this blow-up corresponds to the choice of a phase parameter given that the choice of the spin frame is not unique. More precisely, consider a fixed spin frame $\{ \bmepsilon_{\bmA}{} \}$ at $i$ and $\bmt \in \text{SU}(2,\mathbb{C})$, the transformed spin frame $\bmepsilon_{\bmA}(\bmt) \equiv t_{\bmA}{}^{\bmB} \bmepsilon_{\bmB}{}$ can be extended to an open ball $B_{a}(i)$ in $\mathcal{S}'$ of radius $a$ centred at $i$ by parallel propagation along an $\bmh$-geodesic starting at $i$. If $\rho$ is the affine parameter along the geodesic, then for a fixed $\bmt$, the propagated spin frame can be written as $\bmepsilon_{\bmA}(\rho, \bmt)$. Next, define $\mathcal{M}_{a,\kappa}$, a submanifold of $\mathbb{R} \times \mathbb{R} \times \text{SU}(2,\mathbb{C})$ as
\begin{equation}
    \mathcal{M}_{a,\kappa} = \{ (\tau, \rho, \bmt) \in \mathbb{R} \times \mathbb{R} \times \text{SU}(2,\mathbb{C})| \hspace{1mm} 0 \leq \rho < a, - \frac{\omega}{\kappa} \leq \tau \leq \frac{\omega}{\kappa} \},
    \label{Definiton-M-a-k}
\end{equation}
with the following subsets
\begin{subequations}
    \begin{equation}
        \mathscr{I}^{\pm}_a = \{ (\tau, \rho, \bmt) \in \mathcal{M}_{a,\kappa} | \hspace{1mm} 0 <\rho < a, \tau = \pm \frac{\omega}{\kappa} \}, \qquad \text{past and future null infinity}
    \end{equation}
    \begin{equation}
        \mathcal{I} = \{ (\tau, \rho, \bmt) \in \mathcal{M}_{a,\kappa} | \hspace{1mm} \rho =0, -1 <\tau <1 \}, \qquad \text{the cylinder at spatial infinity}
    \end{equation}
    \begin{equation}
        \mathcal{I}^{\pm} = \{ (\tau, \rho, \bmt) \in \mathcal{M}_{a,\kappa} | \hspace{1mm} \rho =0, \tau = \pm 1 \}, \qquad \text{the critical sets of null infinity}
    \end{equation}
\end{subequations}
and 
\begin{equation}
    \mathcal{I}_0 = \{ (\tau, \rho, \bmt) \in \mathcal{M}_{a,\kappa} | \hspace{1mm} \rho =0, \tau = 0 \}.
\end{equation}

To relate the structures on the fibre bundle to the spacetime manifold $(\tilde{\mathcal{M}}, \tilde{\bmg})$ satisfying \eqref{Vacuum-Einstein-field-eqs}, let $(\mathcal{M}, \bmg)$ denote a smooth conformal extension such that i) $\Theta > 0$ and $\bmg = \Theta^2 \tilde{\bmg}$ on $\tilde{\mathcal{M}}$, ii) $\Theta = 0$ and $d \Theta \neq 0$ on $\mathscr{I}^{\pm}_a$. Now let $\mathcal{N} \subset \mathcal{M}$ denote the domain of influence of $B_{a}(i) \setminus \{ i \}$, then the projection map $\bar{\pi}'$ from $\mathcal{M}_{a,\kappa}$ to $\mathcal{N}$ can be factored as
\begin{equation*}
    \mathcal{M}_{a,\kappa} \xrightarrow{\bar{\pi}'_{1}} \mathcal{M}'_{a,\kappa} \xrightarrow{\bar{\pi}'_{2}} \mathcal{N},
\end{equation*}
where $\mathcal{M}'_{a,\kappa} \equiv \mathcal{M}_{a,\kappa} / \text{U}(1)$ is implied by the action of $\text{U}(1)$ on $\text{SU}(2,\mathbb{C})$. From \eqref{Definiton-M-a-k}, the map $\bar{\pi}'_{1}$ is given by the identity on the $\mathbb{R}\times \mathbb{R}$ component, and, under the identification $\text{SU}(2, \mathbb{C}) = \mathbb{S}^{3}$, by the Hopf fibration on the $\text{SU}(2, \mathbb{C})$ component. In other words, $\bar{\pi}'_{1}$ maps $\mathcal{M}_{a,\kappa}$ onto $\mathbb{R} \times \mathbb{R} \times \mathbb{S}^{2}$.

Finally, note that the spin frames $\bmepsilon_{\bmA}(\rho, \bmt)$ can be extended to the spacetime $\mathcal{M}_{a,\kappa}$ by a certain propagation law along the conformal geodesics orthogonal to the $\mathcal{S}_{a}$, where $\mathcal{S}_{a}$ can be thought of as the initial hypersurface on $\mathcal{M}_{a,\kappa}$, i.e.
\begin{equation*}
    \mathcal{S}_{a} = \{ (\rho, \bmt) \in \mathbb{R} \times \text{SU}(2,\mathbb{C})| \hspace{1mm} 0 \leq \rho < a \}.
\end{equation*}
The propagated spin frames $\bmepsilon_{\bmA}(\tau, \rho, \bmt)$ are determined at any $p \in \mathcal{M}_{a,\kappa} \setminus (\mathcal{I} \cup \mathcal{I}^{+} \cup \mathcal{I}^{-})$ up to a multiplication factor that corresponds to the action of $\text{U}(1)$ on $\text{SU}(\mathcal{M})$. 

\begin{remark}
    Friedrich's formulation involves encoding the F-gauge conditions in the initial data and the properties of the fields appearing in the conformal field equations. For further discussions of Friedrich's formulation and the F-gauge, readers are referred to \cite{Friedrich98, FriedrichKannar00}.
\end{remark}
\begin{remark}
     Fields on $\mathcal{M}_{a,\kappa}$ can be decomposed in terms of complex-valued functions {\em $T_{m}{}^{j}{}_{k}: \text{SU}(2,\mathbb{C}) \to \mathbb{C}$}, closely related to the standard spin-weighted harmonics on $\mathbb{S}^2$ ---see e.g. \cite{FriedrichKannar00}. The analysis of the conformal field equations can be carried out on $\mathcal{M}_{a,\kappa}$ and their solutions can be projected onto $\mathbb{R} \times \mathbb{R} \times \mathbb{S}^{2}$ and used to evaluate BMS asymptotic charges at the critical sets.
\end{remark}
\subsection{The supertranslation asymptotic charges in full GR}
To introduce BMS asymptotic charges at $\mathscr{I}^{\pm}$, let $d^{\bullet}_{abcd}$ denote the rescaled Weyl tensor in the NP-gauge and introduce the spinorial counterpart $d^{\bullet}_{AA'BB'CC'DD'}$ which can be decomposed as follows
\begin{equation}
    d^{\bullet}_{AA'BB'CC'DD'} = - \phi^{\bullet}_{ABCD} \epsilon^{\bullet}_{A'B'} \epsilon^{\bullet}_{C'D'} - \bar{\phi}^{\bullet}_{A'B'C'D'} \epsilon^{\bullet}_{AB} \epsilon^{\bullet}_{CD},
    \label{Decomposition-rescaled-weyl-tensor}
\end{equation}
where $\phi^{\bullet}_{ABCD}$ is a symmetric valence 4 spinor. Given the above, the asymptotic charges associated with smooth functions $f$ on $\mathbb{S}^2$ can be written as
\begin{equation}
    \mathcal{Q}(f;\mathcal{C}) \equiv  \oint_{\mathcal{C}} \bm{\varepsilon}_2 f (\mathcal{P}^{\bullet} - i (*\mathcal{P}^{\bullet}) +\tfrac{1}{2}\sigma^{\bullet ab}N^{\bullet}_{ab}),
    \label{Asymptotic-charges-full-GR}
\end{equation}
where $\mathcal{C}$ is some cross-section of $\mathscr{I}^{\pm}$ and $\bm{\varepsilon}_2$ is its area element, $\sigma^{\bullet ab}$ is the shear tensor, $N^{\bullet}_{ab}$ is the news tensor and 
\begin{subequations}
    \begin{equation*}
        \mathcal{P}^{\bullet} \equiv d^{\bullet}_{abcd} l^a n^b l^c n^d,
    \end{equation*}
    \begin{equation*}
        (*\mathcal{P}^{\bullet}) \equiv (*d^{\bullet})_{abcd} l^a n^b l^c n^d.
    \end{equation*}
\end{subequations}
Using \eqref{Decomposition-rescaled-weyl-tensor} and \eqref{NP-tetrad-in-spinors}, one gets
\begin{equation}
        \mathcal{P}^{\bullet} - i (*\mathcal{P}^{\bullet}) = - 2 \bar{\phi}^{\bullet}_2.
        \label{P-*P-terms}
\end{equation}
Moreover, the term involving $\sigma^{\bullet ab}N^{\bullet}_{ab}$ can be written in terms of the NP-connection coefficients \cite{PenRind84,PenRind86,Stewart91}, whose explicit form depends on whether we are considering the asymptotic charges at $\mathscr{I}^{+}$ or $\mathscr{I}^{-}$. In particular, 
\begin{subequations}
    \begin{equation}
        \sigma^{\bullet ab} N^{\bullet}_{ab}  = 2 \Delta |\sigma^{\bullet}|^2 -|\sigma^{\bullet}|^2 \big(3\mu^{\bullet}+3\bar{\mu}^{\bullet}+\gamma^{\bullet}+\bar{\gamma}^{\bullet}  \big), \qquad \text{on } \mathscr{I}^{+},
    \end{equation}
    \begin{equation}
        \sigma^{\bullet ab}N^{\bullet}_{ab} = 2 \Delta |\lambda^{\bullet}|^2 - |\lambda^{\bullet}|^2 \big(3\rho^{\bullet}+3\bar{\rho}^{\bullet}+\epsilon^{\bullet}+\bar{\epsilon}^{\bullet}  \big), \qquad \text{on } \mathscr{I}^{-}.
    \end{equation}
    \label{The-background-term}
\end{subequations}
Here, $\Delta \equiv n^{a} \nabla^{\bullet}_{a}$ and $\sigma^{\bullet}, \mu^{\bullet}, \gamma^{\bullet}, \lambda^{\bullet},\rho^{\bullet}, \epsilon^{\bullet}$ are the NP-connection coefficients defined as
\begin{subequations}
    \begin{equation}
        \sigma^{\bullet} \equiv -\Gamma^{\bullet}_{\bmzero \bmone'}{}^{\bmone}{}_{\bmzero}, \qquad \mu^{\bullet} \equiv -\Gamma^{\bullet}_{\bmzero \bmone'}{}^{\bmzero}{}_{\bmone}, \qquad \gamma^{\bullet} \equiv \Gamma^{\bullet}_{\bmone \bmone'}{}^{\bmzero}{}_{\bmzero},
    \end{equation}
    \begin{equation}
        \lambda^{\bullet} \equiv \Gamma^{\bullet}_{\bmone \bmzero'}{}^{\bmzero}{}_{\bmone}, \qquad \rho^{\bullet} \equiv -\Gamma^{\bullet}_{\bmone \bmzero'}{}^{\bmone}{}_{\bmzero}, \qquad \epsilon^{\bullet} \equiv \Gamma^{\bullet}_{\bmzero \bmzero'}{}^{\bmzero}{}_{\bmzero}.
    \end{equation}
    \label{NP-connection-ceofficients}
\end{subequations}
In the above, $\bar{\mu}^{\bullet}, \bar{\gamma}^{\bullet}, \bar{\rho}^{\bullet}$ and $\bar{\epsilon}^{\bullet}$ refer to the complex conjugates of $\mu^{\bullet}, \gamma^{\bullet}, \rho^{\bullet}$ and $\epsilon^{\bullet}$, respectively. To evaluate the expression of the charges \eqref{Asymptotic-charges-full-GR} at the critical sets $\mathcal{I}^{\pm}$, one must find a transformation between the NP-gauge frame and the F-gauge frame in full GR. Following \cite{FriedrichKannar00}, a general transformation between a NP-gauge spin frame $\{ \bmepsilon^{\bullet}_{\bmA} \}$ and an F-gauge spin frame $\{ \bmepsilon_{\bmA} \}$ is parameterised by a conformal factor $\theta$ and an $\text{SL}(2,\mathbb{C})$ transformation matrix $\Lambda^{\bmB}{}_{\bmA}$
\begin{equation*}
    \bmepsilon^{\bullet}_{\bmA} = \theta^{-1/2} \Lambda^{\bmB}{}_{\bmA} \bmepsilon_{\bmB},
\end{equation*}
implying transformations for $\bar{\phi}^{\bullet}_{2}$ and the NP-connection coefficients \eqref{NP-connection-ceofficients}. The expressions for these will not be presented here. 

As we are interested in evaluating the expressions of the charges at $\mathcal{I}^{\pm}$, an asymptotic solution for the conformal field equations is analysed, given the initial data prescribed in the previous section. Given the zero-order solution, asymptotic expansions for the conformal factor $\theta$ and the transformation matrices $\Lambda^{\bmB}{}_{\bmA}$ are obtained, following \cite{FriedrichKannar00}. 

If $\phi_{0}, \phi_{1}, \phi_{2}, \phi_{3}, \phi_{4}$ denote the components of the rescaled Weyl tensor in the F-gauge, then the explicit transformation from the NP-gauge to the F-gauge implies
\begin{enumerate}
    \item Contributions to $\mathcal{Q}|_{\mathcal{I}^{\pm}}$ from 
    $\phi_{0}, \phi_{1}, \phi_{3}, \phi_{4}$ are at most $O(\rho)$. 
    \item The background term $\sigma^{\bullet ab} N^{\bullet}_{ab}$ does not contribute to $\mathcal{Q}|_{\mathcal{I}^{\pm}}$ at zero order in $\rho$.
\end{enumerate}
Hence, the asymptotic charges at $\mathcal{I}^{\pm}$ are determined by $f$ and the zero-order solution of $\phi_{2}$, i.e.,
\begin{equation}
    \mathcal{Q}|_{\mathcal{I}^{\pm}} = \mathcal{Q}|_{\mathcal{I}^{\pm}} (f, \phi_{2}^{(0)}).
    \label{Asymptotic-charges-full-GR-2}
\end{equation}
Given that the equation for $\phi_{2}^{(0)}$ is equivalent to the equation for $\psi_{2}$ on Minkowski spacetime, the solution for $\phi_{2}^{(0)}$ will develop a logarithmic singularity at $\mathcal{I}^{\pm}$ unless our initial data satisfy certain regularity conditions. The explicit form of these regularity conditions will be presented in a later article as well as the final expression of $\mathcal{Q}|_{\mathcal{I}^{\pm}}$. The main result is that given initial data that satisfy our regularity conditions, one can show that $\mathcal{Q}|_{\mathcal{I}^{\pm}}$ are fully determined by $\Pi_{3}[\Omega]$. Moreover, if the initial data are chosen to satisfy the regularity conditions, the asymptotic charges $\mathcal{Q}_{l,m}$ associated with a given harmonic $Y_{l,m}$ at $\mathcal{I}^{+}$ and $\mathcal{I}^{-}$ are related by:
\begin{equation}
    \mathcal{Q}_{l,m}|_{\mathcal{I}^{+}}= (-1)^{l} \mathcal{Q}_{l,m}|_{\mathcal{I}^{-}}
\end{equation}

\section{Conclusions}
This article addresses the matching of the asymptotic charges associated with supertranslation symmetries in the context of an initial value problem using Friedrich's formulation of spatial infinity. The results in this paper demonstrate that the zero-order solution of $\phi_{2}$ develops logarithmic singularities at $\mathcal{I}^{\pm}$ given the prescribed initial data in Section \ref{Section:Friedrich's regular initial value problem}. Therefore, $\mathcal{Q}|_{\mathcal{I}^{\pm}}$ are only well-defined if extra regularity conditions are imposed on our initial data. An upcoming article will present the explicit form of these regularity conditions. A significant consequence of this result is that the matching of the BMS asymptotic charges, as defined in this article, is not feasible for generic asymptotically flat spacetimes unless these spacetimes are the development of initial data satisfying certain regularity conditions. 
%%%%%%%%%% Insert bibliography here %%%%%%%%%%%%%%

\bibliographystyle{unsrt}
\bibliography{refs}

\end{document}